\documentclass[aps,12pt]{revtex4}
\usepackage{epsfig}
\usepackage{graphicx}
\usepackage{amsmath,amssymb,color}
\usepackage[english]{babel}
\usepackage{latexsym}
\usepackage{amsfonts}
\usepackage{ulem}

\parskip=\medskipamount

\newcommand{\eq}[1]{(\ref{#1})}
\newcommand{\fig}[1]{Fig.~\ref{#1}}

\newcommand{\be}{\begin{equation}}
\newcommand{\ee}{\end{equation}}

\newcommand\disp{\displaystyle}

\newcommand{\la}{\langle}
\newcommand{\ra}{\rangle}

\begin{document}

\title{From steady-state TASEP model with open boundaries to 1D Ising model at negative fugacity}

\author{Mikhail V. Tamm$^{1,2,*}$, Maxym Dudka$^{3,4}$, Nikita Pospelov$^{5}$,  Gleb Oshanin$^{6}$, Sergei Nechaev$^{7,8}$}

\affiliation{\bigskip
$^1$CUDAN Open Lab, Tallinn University, 10120 Tallinn, Estonia \\ $^2$ Faculty of Physics, Lomonosov Moscow State University, 119992 Moscow, Russia \medskip \\
$^3$Institute of Condensed Matter Physics, National Academy of Sciences of Ukraine, 79011 Lviv, Ukraine \\ $^4$${\mathbb L}^4$ Collaboration $\&$ Doctoral College for the Statistical Physics of Complex Systems, Leipzig-Lorraine-Lviv-Coventry, Europe \medskip \\
$^5$Institute for Advanced Brain Studies, Lomonosov Moscow State University, 119992 Moscow, Russia \medskip \\
$^6$Laboratoire de Physique Th\'eorique de la Mati\`{e}re Condens\'ee (UMR CNRS 7600), Sorbonne Universit\'e, CEDEX 05, 75252 Paris, France \medskip \\
$^7$Interdisciplinary Scientific Center Poncelet (CNRS IRL 2615), 119002 Moscow, Russia \\ $^8$P.N. Lebedev Physical Institute RAS, 119991 Moscow, Russia}

\begin{abstract}

We expose a series of exact mappings between particular cases of four statistical physics models: (i) equilibrium 1D lattice gas with nearest-neighbor repulsion, (ii) (1+1)D combinatorial heap of pieces, (iii) directed random walks on a half-plane, and (iv) 1D totally asymmetric simple exclusion process (TASEP). In particular, we show that generating function of a 1D steady-state TASEP with open boundaries can be interpreted as a quotient of partition functions of 1D hard-core lattice gases with one adsorbing lattice site and negative fugacity. This result is based on the combination of a representation of a steady-state TASEP configurations in terms of (1+1)D heaps of pieces and a theorem of X. Viennot which projects the partition function of (1+1)D heaps of pieces onto that of a {\it single} layer of pieces, which in this case is a 1D hard-core lattice gas.

\bigskip
$^*$ e-mail: thumm.m@gmail.com

\end{abstract}

\maketitle

\section{Introduction}

The Totally Asymmetric Simple Exclusion Process (TASEP) is a stochastic process involving a concentration of hard-core particles which perform random, totally directed walks on a regular one-dimensional lattice subject to the constraint that each lattice site may sustain at most one particle. More specifically, updating rules are defined as follows: each particle attempts to jump to the neighboring lattice site on its right with a given rate, which can be chosen as 1 without any loss of generality, and the jump is actually fulfilled if and only if the target site is empty at this time instant, otherwise the jump is rejected. The jumps in the opposite direction are forbidden. The model possesses a particle-hole symmetry, i.e., it is symmetric with respect to a simultaneous replacement of particles with holes and vice versa and an inversion of the direction of motion. A detailed introduction, definitions and a review of important results obtained for this process can be found in \cite{derrida98}.

\begin{figure}[h]
\epsfig{file=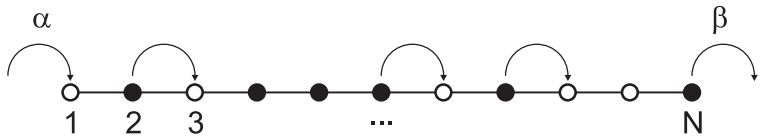,width=10cm}
\caption{TASEP on a chain containing $N=11$ sites. The chain is attached to a reservoir of particles at $j = 1$ which ``adds'' particles to the system with the constant rate $\alpha$ whenever this site is empty. The particles are removed from the system with the constant removal rate $\beta$ at the site $j = N$.}
\label{f:03}
\end{figure}

TASEP on a finite chain of $N$ sites attains a non-equilibrium steady state which depends on the boundary conditions used. The two most typical choices of the latter are: (a) periodic boundary conditions, i.e., the chain forms a ring so that the number of particles initially introduced into the system is conserved, and (b) the chain has open boundaries: on the left extremity it is attached to an infinite reservoir of particles maintained at a constant chemical potential, while on the right extremity there is another infinite reservoir, which also has a constant chemical potential, smaller than the one on the left. Consequently, the particles are injected into the system on the left boundary at a constant rate $\alpha$ provided that this leftmost site (with $j = 1$) is empty at this time instant and, whenever they reach the rightmost site $j = N$ they are removed with a constant rate $\beta$. A sketch of such a model is presented in \fig{f:03}. Note that in the former case the steady state is very simple: all configurations respecting the conservation of the number of particles are equiprobable (see, e.g., \cite{krapivskybook}). On contrary, in the latter case the system evolves towards an out-of-equilibrium steady-state with a non-trivial particle density distribution, which has been determined via a matrix ansatz in \cite{DEHP}. Concurrently, combinatorial interpretations of the steady-state weights of different configurations have been obtained earlier in terms of pairs of paths (for $\alpha=\beta=1$) in \cite{Shapiro82}, and also in terms of weighted permutation tableaux \cite{Corteel07} and weighted binary trees \cite{Viennot07}.

In this paper we demonstrate that the generating function of a steady-state TASEP with open boundaries can be represented in terms of partition functions of a 1D hard-core lattice gas at a negative fugacity (i.e., at a purely imaginary chemical potential) and with one adsorbing lattice site. To show that we exploit a bijection (first discussed in \cite{haug}) between the TASEP and the so-called ``heaps of pieces'' (HP) model \cite{viennot-rev}. Further on, we take advantage of a theorem which links the HP model and a certain model of a lattice gas of hard-core objects, first established by X. Viennot in \cite{viennot-rev}.

The paper is organized as follows. In Sec. \ref{a} we remind the matrix ansatz for the TASEP with open boundaries. In Sec. \ref{b} we describe the HP model, present the definitions of the so-called Mikado ordering and of the \L{}ukasiewicz paths and establish a connection between the TASEP and the HP model via a direct enumeration of the Mikado orderings. Next, in Sec. \ref{c} we recall the Viennot theorem and eventually show that the generating function of the steady-state TASEP on a chain with open boundaries can be represented in terms of partition functions of a 1D hard-core lattice gas with a negative fugacity, one adsorbing site and a special kind of boundary conditions. Finally, in the Discussion we present a brief summary of the results and outline some open questions. In Appendix I we recall the approach to an enumeration of (1+1)D heaps based on the geometric group theory, and in Appendix II we outline the connection between \L{}ukasiewicz paths introduced in the main text, the Brownian excursions and the Young tableaux.

\section{Matrix Ansatz for the TASEP on a chain with open boundaries}
\label{a}

We start by recalling the matrix ansatz for the steady state of the TASEP model on an $N$-site chain with constant entrance, $\alpha$, and exit, $\beta$, rates \cite{DEHP}. To this end, we first introduce two formal operators $D$ and $E$, which satisfy the relation
\be
DE = D+E \,,
\label{eq:10}
\ee
and two vectors $\la {\bf V}_{out}|$ and $|{\bf V}_{in}\ra$, such that
\be
D|{\bf V}_{in}\ra = \beta^{-1}|{\bf V}_{in}\ra; \quad \la {\bf V}_{out}|E = \alpha^{-1} \la {\bf V}_{out}|.
\label{eq:11}
\ee
Then, the probability of observing any given configuration in the steady state is proportional to a matrix element of the form $\la {\bf V}_{out}|...|{\bf V}_{in}\ra$, where in place of dots one should insert a sequence of $N$ operators $D$ and $E$, with $D$ and $E$ corresponding to occupied and empty sites, respectively. To write this down in more formal terms, introduce occupation numbers of the sites of a chain, $\sigma_i$, $(1\le i\le N$), such that $\sigma_i=1$ if the $i$-th site is occupied by a particle and $\sigma_i=0$, otherwise, and define the probability $P(\sigma|t)$ to have a set of occupation numbers, $\sigma = \{\sigma_1,\sigma_2,...,\sigma_N\}$ at time instant $t$. In the steady state,
\be
\frac{d}{dt}P(\sigma|t)=0 \,.
\label{eq:12}
\ee
Dropping the argument $t$, one writes next the probability $P(\sigma)$ in the steady state as follows
\be
P(\sigma)=\frac{1}{Z_N(\alpha,\beta)}f(\sigma),
\label{eq:13}
\ee
where the weight $f(\sigma)$ of the configuration $\{\sigma_1,\sigma_2,...,\sigma_N\}$ is
\be
f(\sigma)=\left<{\bf V}_{out}\right|\prod_{i=1}^N\left(\sigma_i D+(1-\sigma_i)E \right) \left|{\bf V}_{in}\right>.
\label{eq:14}
\ee
For example, the weight $f(\{\sigma\})$ of the configuration shown in \fig{f:03} is $f=\la {\bf V}_{out}| E D E D D D E D E E D |{\bf V}_{in}\ra$. The normalization factor $Z_N$, which is often called the (non-equilibrium) partition function, is given by
\be
Z_N(\alpha,\beta) = \sum_{\tau_1=\{0,1\}}...\sum_{\tau_N=\{0,1\}} f(\tau_1,\tau_2,...,\tau_N) = \left<{\bf V}_{out}\right|\left(D+E \right)^N \left|{\bf V}_{in}\right> .
\label{eq:15}
\ee

Except for some particular values of $\alpha$ and $\beta$, the algebra defined by \eq{eq:10} and \eq{eq:11} has no finite-dimensional representations. However, there exist many infinite-dimensional ones, among which the most interesting for us is the one constructed in the following way. Take
\be
\la {\bf V}_{out}|= (1,\alpha^{-1},\alpha^{-2}, \alpha^{-3},\dots),\quad \la {\bf V}_{in}|=(1,0, 0, 0, \dots)
\label{boundaries}
\ee
and choose the infinite-dimensional matrices $D$ and $E$ in the form
\be
D = \left(
\begin{array}{cccccccc}
\frac{1}{\beta} & \frac{1}{\beta} & \frac{1}{\beta} & \frac{1}{\beta} & \frac{1}{\beta} & \frac{1}{\beta} & \ldots \\
0 & 1 & 1 & 1 & 1 & 1 & \ldots \\
0 & 0 & 1 & 1 & 1 & 1 & \ldots \\
0 & 0 & 0 & 1 & 1 & 1 & \ldots \\
0 & 0 & 0 & 0 & 1 & 1 & \ldots \\
0 & 0 & 0 & 0 & 0 & 1 & \ldots \\
\vdots & \vdots & \vdots & \vdots & \vdots & \vdots & \ddots
\end{array}\right); \quad
E = \left(
\begin{array}{cccccccc}
0 & 0 & 0 & 0 & 0 & 0 & \ldots \\
1 & 0 & 0 & 0 & 0 & 0 & \ldots \\
0 & 1 & 0 & 0 & 0 & 0 & \ldots \\
0 & 0 & 1 & 0 & 0 & 0 & \ldots \\
0 & 0 & 0 & 1 & 0 & 0 & \ldots \\
0 & 0 & 0 & 0 & 1 & 0 & \ldots \\
\vdots & \vdots & \vdots & \vdots & \vdots & \vdots & \ddots
\end{array}\right).
\label{eq:17}
\ee
Then it can be checked directly that both conditions \eq{eq:10} and \eq{eq:11} are fulfilled. In what follows we show that the partition function \eq{eq:15} can be interpreted as a result of a direct enumeration of weighted heaps of pieces in (1+1)D for some special choice of weights of heaps.

\section{Connection between the TASEP and the HP model}
\label{b}

\subsection{Definition of the HP model}

A heap of pieces is a collection of elements which are piled together along the vertical axis. If two elements intersect or touch each other in their horizontal projections, then the resulting heap depends on the order in which these two were placed: the element which is placed second is above the element placed first. Such rules resemble the famous {\it tetris} computer game, in which pieces of various shapes are dropped down along vertical direction until they hit the already deposited elements.

A heap has a base -- a set of all possible positions in the direction orthogonal to the vertical axis. Bases of various forms can be considered, including lattices in various dimensions, and, more generally, for any fixed graphs. In turn, shapes of pieces can also be different, as well as rules of their interactions. Apparently, the concept of a heap of pieces has been first proposed in 1969 in the work of P. Cartier and D. Foata \cite{cartier} in which they considered monoids generated by some alphabet with special commutation relations. Variety of models, as well as new combinatorial results and their links with the statistical physics were reviewed in \cite{viennot-rev}. The (1+1)D HP model on square and triangular lattices have been exhaustively studied in the literature and played the role of a testing ground for several approaches -- from purely combinatorial \cite{viennot-rev, betrema, bousquet3}, to the ones based on the diagonalization of the spatial transfer matrix and the Bethe Ansatz computations \cite{hakim1,dhar1,dhar2, dhar}.

Apart from the enumeration of growing heaps, some other problems in pure mathematics and in mathematical physics are connected to the HP model. For example, various aspects \cite{hakim1, viennot1, bousquet1, bousquet2} of the enumerative combinatorics of partitions are related to a growth of (1+1)D HP. In \cite{vershik} the statistics of growing heaps has been linked to the statistics of two-dimensional growing braids, in \cite{anim-math} the general asymptotic theory of directed two-dimensional lattice paths in half-planes and quarter-planes has been reviewed.

\begin{figure}[h]
\epsfig{file=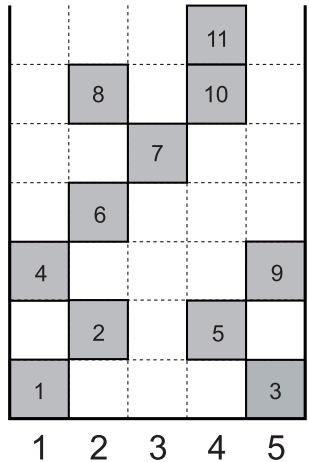, width=4cm}
\caption{A sketch of a particular configuration of a heap with $N=11$ pieces in a bounding box of size $n=5$.}
\label{f:sample}
\end{figure}

One of the main questions in the study of the HP problem is the analysis of an asymptotic behavior of the partition function
\be
Z_N \sim N^{\theta}\Lambda^N,
\label{Lambda-def}
\ee
which enumerates all allowed distinct configurations of $N$-particle heaps ($N\to \infty$) over a given base graph. In case when the base is a D-dimensional lattice of a linear extent $n$, the critical exponent, $\theta$, is universal and depends only on the space dimensionality, while $\Lambda$ depends on $n$, on the lattice geometry, the shape of the pieces, and also on the way how the interactions between them are defined. Here we discuss the heaps of {\it square} pieces which cannot touch each other by their {\it side} faces, top and bottom faces are allowed to touch (see \fig{f:sample} for a typical configuration of such a heap).

One can imagine a heap of pieces as in \fig{f:sample} resulting from some deposition process with pieces falling down from $y=+\infty$ until they reach the lowest possible position respecting the constraint that no pieces have common vertical faces. Numbers inside the falling blocks designate sequential discrete moments of time at which corresponding piece is added to a heap. However, there is an important distinction between the enumeration of configurations in HP and the enumeration of different states in a sequence of falling blocks. In the HP problem, as described above, we are interested in the total number of possible configurations, which respect the rules of a heap's formation (in this case -- the absence of touching side faces of the squares). Thus we imply that all allowed heaps have equal weights. On the other hand, in a deposition problem, although the total set of allowed heaps is the same, there is no such equiprobability: some heaps are obtained more often than others. Let us therefore stress that in what follows we consider just the {\it combinatorial} HP problem rather than the dynamical deposition one.

There exists a connection (first revealed in \cite{haug}) between the partition function of the (1+1)D heap of square pieces with no touching vertical faces and the partition function of a steady-state of the TASEP with open boundary conditions. We describe this connection in the subsequent parts of this section.

\subsection{Mikado ordering and transfer matrix approach to the HP model}

Let us outline the computations of the partition function $Z_N(n)$ of the (1+1)D heap of square pieces of the type shown in \fig{f:sample}. First, we introduce a unique enumeration of heaps, then we show that, given that enumeration, it is possible to write a transfer matrix equation for $Z_N(n)$. Finally, we notice that this equation resembles the one for the partition function of the TASEP with open boundaries.

\begin{figure}[ht]
\epsfig{file=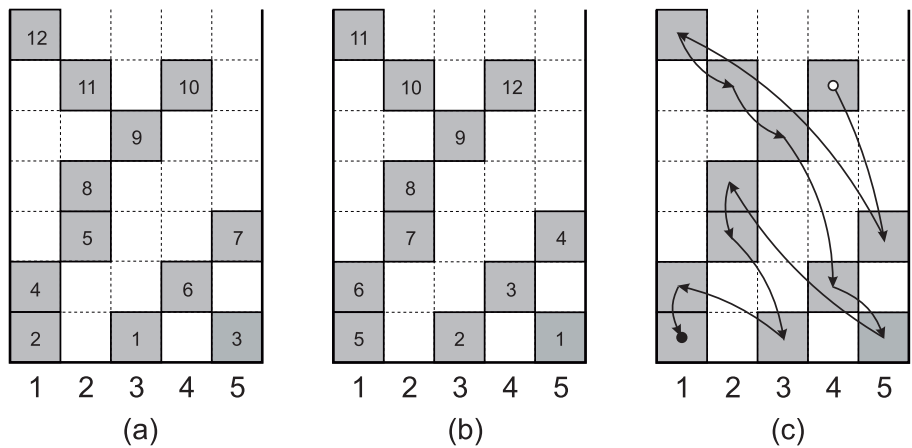, width=12cm}
\caption{Particular realization of a heap. The heap in (a) is obtained by the sequential dropping of bricks and spells as a word $W_a=g_3g_1g_5g_1g_2g_4g_5g_2g_3g_4g_2g_1$; the same heap in (b) is obtained by the sequential dropping of bricks corresponding to another sequence $W_b=g_5g_3g_4g_5g_1g_1g_2g_2g_3g_2g_1g_4$; (c) the unique ``Mikado ordering'' of pieces, see the text for description.}
\label{f:01}
\end{figure}

\begin{figure}[ht]
\epsfig{file=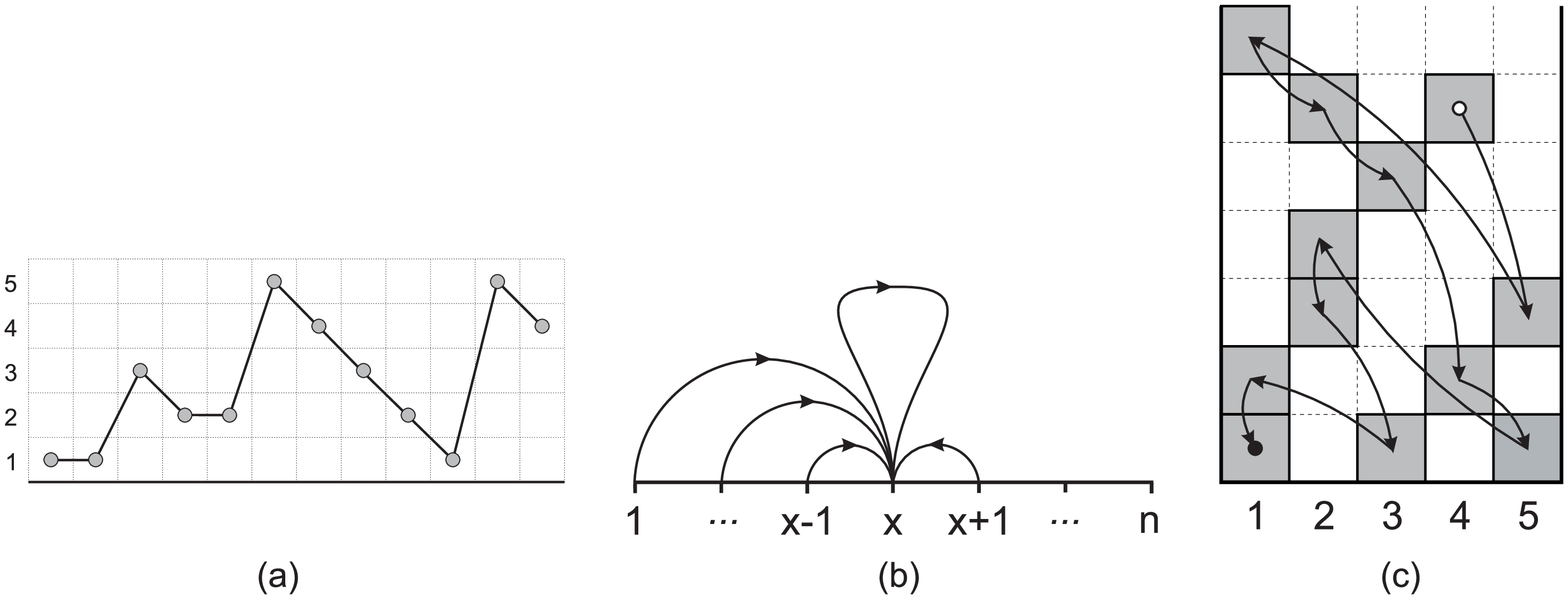, width=16cm}
\caption{(a) \L{}ukasiewicz path corresponding to the Mikado ordering shown in figure (c). (b) the allowed steps of the \L{}ukasiewicz walk: if $g_x$ is followed by $g_y$, $y$ cannot be larger than $x+1$.}
\label{f:02}
\end{figure}

As we noticed above, each heap can be thought of as a result of some deposition process. However, as it is shown in \fig{f:01}a,b different deposition sequences (depicted by numbers inside the pieces) can lead to a same geometrical heap. It is thus essential to define a rule allowing to enumerate pieces of a heap in a unique way. To do that, note that each heap has at least one piece which satisfies the following two conditions: (i) if it is removed the remaining part is itself a valid heap, and (ii) if it is redeposited (i.e., deposited from above into the same column), the original heap is recovered. We call the set of such ``allowed'' pieces the ``roof'' of a heap. In order to enumerate pieces in a unique way we proceed as follows. We fix the position of the {\it rightmost} element in the roof of the heap and remove this piece. The remaining heap has one piece less, and it itself has an updated roof, so one can repeat the removal procedure until the heap gets empty. As a result, e.g., for the heap shown in \fig{f:02}c, we get the following order of removed pieces
\be
\overleftarrow{W}=g_4\,g_5\,g_1\,g_2\,g_3\,g_4\,g_5\,g_2\,g_2\,g_3\,g_1\,g_1,
\label{eq:04}
\ee
where we use letters (``generators'', in notations of Appendix 1 where a discussion of the underlying group-theoretical construction is outlined) $g_i$ to denote pieces in the $i$-th column. We call such an enumeration procedure the {\it Mikado ordering} because it resembles the famous Mikado game, the goal of which consists in a sequential removal of sticks from a pile, one-by-one, without disturbing the rest of a pile. By construction, each heap has a unique Mikado ordering. Moreover, inverse Mikado ordering
\be
\overrightarrow{W}=g_1\,g_1\,g_3\,g_2\,g_2\,g_5\,g_4\,g_3\,g_2\,g_1\,g_5\,g_4,
\label{eq:04a}
\ee
corresponds to a specific sequence of deposition of pieces that results in a heap shown in \fig{f:01}c. This proves that each Mikado ordering produces a unique heap, i.e., there is a one-to-one correspondence between heaps and their Mikado orderings. So, given a particular configuration of a heap (no matter how it is created), we associate with it a unique sequence of letters constructed according to Mikado rule.

It is natural to represent the Mikado orderings by graphs as shown in \fig{f:02}, where the horizontal coordinate is the position of a piece in the Mikado ordering (ordered from right to left as in \eq{eq:04}) and the vertical coordinate is the coordinate of a piece (index of the generator $g$). One can interpret such graphs as some discrete-space walks on the $x=1,\dots, n$ interval. On each step a walker either goes up making an arbitrary number of steps, or stays at the same position, or goes one step down. Paths satisfying these conditions are known in the literature \cite{luk1,luk2} as the \L{}ukasiewicz paths. Clearly, there is one-to-one correspondence between such paths and the Mikado-ordered HPs. Interestingly, there exists a mapping between the \L{}ukasiewicz paths, the standard Dyck paths (discrete one-dimensional directed walks for which only increments of $\pm 1$ are allowed) and the Young tableaux, we discuss this connection in  Appendix 2.

Now, it is possible to calculate the number of Mikado orderings (and thus, the total number of heaps) as follows. Let $Z_N(x,x_0|n)$ be a total number of heaps with the Mikado ordering of pieces starting with a piece positioned at $x$ and ending with a piece positioned at $x_0$ ($1\le x,x_0 \le n$). The function $Z_N(x,x_0|n)$ satisfies the recursion scheme of the form
\be
\left\{\begin{array}{l}
\disp Z_{N+1}(x,x_0|n) = \sum_{x'=1}^{x+1} Z_N(x',x_0|n),\;\; x=1,\dots,n; \medskip \\
Z_{N=0}(x,x_0|n) = \delta_{x,x_0}
\end{array} \right.
\label{eq:05}
\ee
Indeed, the Mikado ordering dictates that on each step one takes the rightmost piece off the roof of the heap. Thus, if at sequential time moments the pieces are removed at positions $x$ and $x'$, respectively, then either $x-x' > 1$ (both pieces belong to the roof at the initial step and $x$ is to the right of $x'$, so it is removed first) or $|x-x'| \leq 1$ (piece in position $x$ originally blocks the piece in position $x'$, but it gets released after piece at $x$ is removed). It is easy to verify (see, e.g., \cite{haug}) that this constraint is sufficient, i.e., that any sequence of pieces respecting the rule $x_{i} \geq (x_{i-1} -1)$ for all $i=1,\dots,N$ can be obtained as a valid Mikado ordering (note, however, that the similar statement is not true for heaps of pieces in higher dimensions \cite{npt20}). The allowed sequences of pieces is schematically depicted in \fig{f:02}c (see the Appendix 1 for more details).

It is convenient to rewrite the recursion \eq{eq:05} in a matrix form as follows
\be
Z_N(x,x_0|n) = \la {\bf X}_{out}\,|\, T^N(n)\,|\, {\bf X}_{in} \ra; \quad {\bf X}_{in}=(\overbrace{0,...,0,1}^{x_0},0,...,0)^{\top}, \; {\bf X}_{out}=(\overbrace{0,...,0,1}^{x},0,...,0),
\label{eq:07}
\ee
where the transfer matrix $T(n)$ reads
\be
T(n)=\left(
\begin{array}{cccccccc}
1 & 1 & 1 & 1 & \ldots & 1 \\
1 & 1 & 1 & 1 & \ldots & 1 \\
0 & 1 & 1 & 1 & \ldots & 1 \\
0 & 0 & 1 & 1 & \ldots & 1 \\
\vdots & \vdots & \vdots & \ddots & \ddots & \vdots \\
0 & 0 & 0 & 0 & \ldots & 1
\end{array}\right);
\label{eq:06}
\ee
while the partition function enumerating all possible heaps is given by
\be
Z_N(n) = \la {\bf Y}_{out}\,|\, T^N(n)\,|\, {\bf Y}_{in} \ra; \qquad {\bf Y}_{in}= (1,1,1,...,1,1)^{\top}, \quad {\bf Y}_{out} = (1,1,1,1,...,1).
\label{eq:08}
\ee
Thus, the growth rate $\Lambda(n)$ defined by \eq{Lambda-def} is determined by the largest eigenvalue of the transfer matrix \eq{eq:06}. The corresponding computation has been repeatedly discussed in the literature (see, e.g., \cite{vershik}), and $\Lambda(n)$ is given by
\be
\Lambda(n) = 4\cos^2\frac{\pi}{n+1}\bigg|_{n\gg 1} \approx 4-\frac{4\pi^2}{n^2}.
\label{eq:09}
\ee
In particular, in a large bounding box of base $n\gg 1$ the growth rate is saturated at the value $\lambda_{\infty} = \lim_{n\to\infty}\Lambda(n) = 4$.

Now, for our purposes it is essential to notice a striking similarity between Eq. \eq{eq:15} and Eqs. \eq{eq:07} and \eq{eq:08}. Indeed, in the limit $n \to \infty$ the transfer matrix \eq{eq:06} coincides with the matrix $(D+E)$, given by \eq{eq:17} for the case of $\beta = 1$. Thus, for $n \to \infty$ one gets
\be
Z_N^{\text{TASEP}} (\alpha, \beta = 1) = \alpha \lim_{n\to\infty} \sum_{y=1}^{n} \disp \alpha^{-y} Z_{N}^{\text{HP}}(x=1|y),
\label{beta1}
\ee
representing the partition function of the TASEP with open boundaries as a weighted sum over partition functions of heaps of pieces with the topmost piece at $x=1$ (such heaps are called ``pyramids'' in Viennot's notations \cite{viennot-rev}), and we took into account the particular forms of $|{\bf V}_{in}\ra$, $\la {\bf V}_{out}|$ given by \eq{boundaries} to arrive at the formula \eq{beta1}.

\begin{figure}[ht]
\epsfig{file=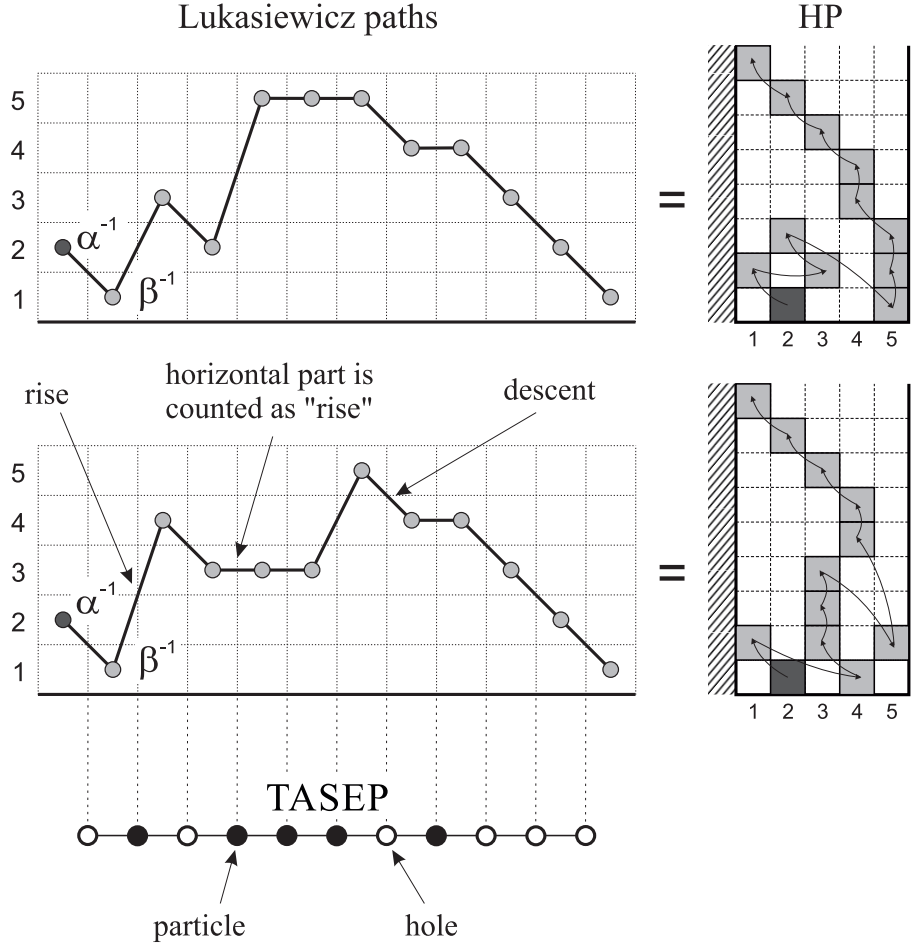, width=12cm}
\caption{Two examples of HP pyramids (right) and their corresponding \L{}ukasiewicz paths (left). Note that, according to the mapping introduced in the text, these two configurations correspond to the same TASEP configuration shown below. If the ``\L{}ukasiewicz walker'' touches the bottom line $x=1$, it gets the weight $\beta^{-1}$ and the very first step, $g_{x_0}$, carries the weight $\alpha^{1-x_0}$. The last step is always at the position $x_N=1$.}
\label{f:05}
\end{figure}

\subsection{Weighted \L{}ukasiewicz paths and TASEP-HP analogy for $\beta \neq 1$}

It is easy to generalize \eq{beta1} to the case of arbitrary $\beta$, one just needs to assign an additional weight $\beta^{-1}$ to a heap every time when there appears a piece with coordinate $x=1$. One can rationalize this by considering an adsorbing vertical wall at $x=0$, so that the pieces in the leftmost column acquire an additional energy compared to the pieces in other columns. In the \L{}ukasiewicz path interpretation (see \fig{f:05}) it means that each path acquires the weight $\beta^{-1}$ every time when it touches the horizontal axis. This problem can be reinterpreted as an adsorption of an ideal polymer at a point-like potential well \cite{grosberg-khokhlov} in 1D. Similar weighted sums over random walk trajectories arise in the context of wetting \cite{wetting}, or path-counting on regular graphs with a defect \cite{tnk, nvt17}. As a result, one gets the following mapping
\be
Z_N^{\text{TASEP}} (\alpha, \beta) = \lim_{n\to\infty}\la {\bf V}_{out}|T_{\beta}(n)^N|{\bf V}_{in} \ra = \alpha \beta \sum_{\text{all~pyramids~of~size~$N$}} \alpha^{-y} \beta^{-(\#(x=1))},
\label{mapping_main}
\ee
where the summation runs over all configurations of pyramids of $N$ pieces, $y$ is the coordinate of the last piece in the Mikado ordering (i.e. the leftmost piece in the lowest layer), and $\#(x=1)$ is the number of pieces with the coordinate $x=1$. For example, both pyramids shown in \fig{f:05} have weight $\alpha^{-1} \beta^{-1}$ because the coordinate of the leftmost piece in the lowest layer is $y=2$ and there are 2 pieces with coordinate equal to 1.

Note that despite this mapping, {\it there is no one-to-one correspondence} between TASEP configurations and heap configurations. Indeed, while the matrix $E$ (corresponding to an empty site in the TASEP setting) can be identified with a descending step of the corresponding \L{}ukasiewicz path, the matrix $D$ corresponds to the summation over all permitted horizontal or ascending steps in the path. Thus, the weight of a given $N$-particle TASEP configuration can be calculated as a weighted sum within the HP model according to the following rules:
\begin{enumerate}
\item[(a)] Summation over HP configurations runs over all possible Mikado ordered sequences with $N+1$ pieces, in which the first piece is $g_1$, any sequence $g_i g_k$ with $k\le i$ corresponds to a particle, and a sequence $g_i g_{i+1}$ corresponds to a hole at the corresponding position of the TASEP configuration,
\item[(b)] The first letter, $g_y$, in the normally ordered word carries a weight $\alpha^{-y}$,
\item[(c)] Each generator $g_1$ carries a weight $\beta^{-1}$,
\item[(d)] Weight of all other generators is $1$,
\item[(e)] In order to obtain standard form of the weight one should multiply the result by $\alpha \beta$. However, since all weights are defined up to a common multiplicative content, this last step bears no additional meaning and is done only for the purposes of comparison with the conventional formulae \cite{DEHP}.
\end{enumerate}

\subsection{Generating function of the stationary TASEP via enumeration of weighted heaps}

Given the specific form of the transfer matrix
\be
T_{\beta}(n)=\left(\begin{array}{cccccc}
\frac{1}{\beta} & \frac{1}{\beta} & \frac{1}{\beta} & \ldots & \frac{1}{\beta} & \frac{1}{\beta} \\
1 & 1 & 1 & \ldots & 1 & 1 \\
0 & 1 & 1 & \ldots & 1 & 1 \\
0 & 0 & 1 & \ldots& 1 & 1 \\
\vdots & \vdots & \vdots & \ddots & \vdots & \vdots \\
0 & 0 & 0 & \ldots & 1 & 1
\end{array}\right),
\label{eq:19}
\ee
it is possible to calculate the right-hand-side of \eq{mapping_main} exactly. The result is, of course, known \cite{DEHP} but it is instructive: (i) to provide a calculation of the matrix element $\la {\bf V}_{out}|T_{\beta}(n)^N|{\bf V}_{in} \ra$ for arbitrary $n$ and (ii) to discuss the interpretation of the well-known stationary TASEP phases in terms of the HP model and the \L{}ukasiewicz paths.

Consider vector ${\bf Z}_N = (Z_N(1), Z_N(2),...,Z_N(n))^{\top}$ defined by recurrence relation
\be
{\bf Z}_{N+1} = T_{\beta}(n) {\bf Z}_N; \quad {\bf Z}_{N=0}={\bf V}_{in} =(1,0,...,0,0)^{\top}
\label{eq:20}
\ee
and introduce a generating function ${W}(s) \equiv (W(s,1),W(s,2),...W(s,n))^{\top} = \sum_{N=0}^{\infty} {\bf Z}_N s^N$. Then
\be
\frac{1}{s}({\bf W}(s) - {\bf Z}_0) = T_{\beta}(n) {\bf W}(s);\;\;\; {\bf W}(s) = - \left(s T_{\beta}(n) - I \right)^{-1} {\bf Z}_0,
\label{eq:21}
\ee
where $I$ is the identity matrix. The elements of vector ${\bf W}(s)$ can be obtained as
\be
W(s,k) = \frac{\det B(k)}{\det (T_{\beta}(n) - \frac{1}{s} I)}=\frac{v_{n,k}}{u_n},
\label{eq:22}
\ee
where the matrix $B(k)$ is obtained from $ (T_{\beta}(n) - \frac{1}{s} I)$ by replacing the $k$-th column with $(-1/s,0,...,0,0)^{\top}$:
\be
B(k) = \left(\begin{array}{cccccc}
1/\beta -1/s & 1/\beta &\ldots &{\bf -1/s} &\ldots &1/\beta
\medskip \\ 1 & 1 -1/s & \ldots& {\bf 0} & \ldots &1 \medskip \\
0 & 1 & \ldots& {\bf 0} & \ldots & 1 \\
0 & 0 & \ldots& {\bf 0} & \ldots & 1 \medskip \\
\vdots & \vdots & \vdots & \vdots & \ddots & \vdots \medskip \\
0 & 0 & \ldots & {\bf 0} & \ldots & 1-1/s
\end{array}\right),
\label{eq:23}
\ee
and $v_{n,k}$ and $u_n$ are the short-hand notations for the numerator and denominator of \eq{eq:22}, respectively. The denominator $u_n$ satisfies, with respect to $n$, the following recurrence relations
\be
\begin{cases} u_{n+2}=-\frac{1}{s}u_{n+1}-\frac{1}{s}u_n; \medskip \\ u_0=1; \medskip \\ u_1=\frac{1}{\beta}-\frac{1}{s}.
\end{cases}
\label{eq:25}
\ee
The solution of \eq{eq:25} has a form
\be
u_n = C_1 p_1^n + C_2 p_2^n,
\label{eq:26}
\ee
where $p_1$ and $p_2$ are the roots of the quadratic equation
\be
p^2 =-\frac{1}{s} p - \frac{1}{s}
\label{eq:27}
\ee
and $C_1$ and $C_2$ are determined from the initial conditions $u_0 = C_1 + C_2$, $u_1 = C_1 p_1 +C_2 p_2$. After some algebra one gets
\be
u_n = u_n(s,\beta)= \frac{\frac{1}{\beta}-\frac{1}{s}-p_2}{p_1-p_2}\,p_1^n- \frac{\frac{1}{\beta}-\frac{1}{s}-p_1}{p_1-p_2}\,p_2^n =\frac{s}{\sqrt{1-4s}} \left(\left(p_1 + \frac{1}{\beta}\right) p_1^n - \left(p_2 + \frac{1}{\beta}\right) p_2^n \right),
\label{eq:29}
\ee
where
\be
p_{1,2}=\frac{-1\pm\sqrt{1-4s}}{2s}.
\label{p12}
\ee
In turn, the determinants in the numerator of \eq{eq:22}, $v_{n,k}=\det B(k)$, can be expressed as:
\be
v_{n,k}(s)= \frac{(-1)^k}{s} u_{n-k}(s,\beta=1).
\label{eq:31}
\ee
Introduce now a generating function
\be
\Xi_n (s,\alpha, \beta) = \sum_{N=0}^{\infty} \la {\bf V}_{out}|T_{\beta}(n)^N|{\bf V}_{in} \ra s^N = \sum_{k=1}^n W(s,k) \alpha^{-k+1},
\label{Xi}
\ee
and substitute eqs. \eq{eq:22}, \eq{eq:29} and \eq{eq:31} into $\Xi_n (s,\alpha, \beta)$ to get
\be
\Xi_n (s,\alpha, \beta)= -\frac{1}{s} \frac{(p_1+1)\dfrac{p_1^n - (-\alpha)^{-n}}{p_1+\alpha^{-1}}-(p_2+1)\dfrac{p_2^n - (-\alpha)^{-n}}{p_2+\alpha^{-1}}}{(p_1+\beta^{-1})p_1^n - (p_2+\beta^{-1})p_2^n}.
\label{Xi_n}
\ee
In the vicinity of $s=0$ \eq{Xi_n} has a well-defined limit for $n \to \infty$
\be
\begin{array}{rll}
\Xi (s,\alpha, \beta) &= &\displaystyle \sum_{N=0}^{\infty} Z_N^{\text{TASEP}} (\alpha, \beta) s^N = \lim_{n \to \infty} \Xi_n (s,\alpha, \beta)= -\frac{1}{s}\frac{p_2+1}{(p_2+\alpha^{-1})(p_2+\beta^{-1})} = \medskip \\
&= & \displaystyle 2\frac{\sqrt{1-4s}+1-2s}{\left(\sqrt{1-4s}+1-2s\alpha^{-1}\right)\left(\sqrt{1-4s}+1-2s\beta^{-1}\right)},
\end{array}
\label{TASEP_Xi}
\ee
which generates partition functions of stationary TASEP. Note the $\alpha \leftrightarrow \beta$ symmetry arises in the $n\to \infty$ limit \eq{TASEP_Xi}, while the expression \eq{Xi_n} does not have this symmetry for any finite $n$ (indeed, it is a polynomial in $\alpha^{-1}$ but an infinite series in $\beta^{-1}$). The large--$N$ behavior of the partition function $Z_N^{\text{TASEP}} (\alpha, \beta)$, and, in particular, the stationary flow,
\be
I = \lim_{N \to \infty} \frac{\la {\bf V}_{out}|T^{N-1}|{\bf V}_{in}\ra }{\la {\bf V}_{out}|T^N|{\bf V}_{in}\ra } = \lim_{N \to \infty} \frac{Z_{N-1} (\alpha, \beta)}{Z_N (\alpha, \beta)},
\ee
is controlled by the smallest (in terms of absolute value) singularity of $\Xi (s,\alpha, \beta)$. Depending on the particular values of $\alpha$ and $\beta$ it could be:
\begin{itemize}
\item[(i)] the square root singularity, $I^* = s_1 = 1/4$, corresponding to the maximal flow phase,
\item[(ii)] the pole $I^{**} = s_2 (\beta)= \beta (1-\beta)$ corresponding to the high density phase,
\item[(iii)] the pole $I^{***} = s_3 (\alpha)= \alpha(1-\alpha)$ corresponding to the low density phase.
\end{itemize}
The transition between these phases occurs at
\be
\begin{array}{rcl}
s_1 = s_2 (\beta) & \rightarrow & \beta = 1/2; \medskip \\
s_1 = s_3 (\alpha) & \rightarrow & \alpha = 1/2; \medskip \\
s_2 (\beta) = s_3(\alpha) & \rightarrow & \beta = \alpha;
\end{array}
\label{eq:derrida}
\ee
in full agreement with \cite{DEHP}.

It is instructive to discuss the interpretation of the TASEP phase transitions \eq{eq:derrida} in terms of the \L{}ukasiewicz paths. The three phases of the stationary TASEP described above (maximal flow, high density and low density) correspond to situations in which typical \L{}ukasiewicz paths are: (i) freely diffusing, (ii) pinned to the absorbing wall, and (iii) fully elongated, respectively. The transition between the diffusive and pinned states indeed is known to occur at a pinning weight $\beta^{-1} = 2$ (see, e.g., \cite{tnk}). In \cite{krug94} it was shown that the path confined between two adsorbing walls with pinning weights $\beta^{-1}, \alpha^{-1}$ is analogous to the TASEP with open boundaries. Here, instead of adsorption to the second wall we have an elongated phase of \L{}ukasiewicz paths, which can be thought of as a result of paths' stretching in external field acting on the first link of the path. The transition between this force-induced phase and the adsorbed phase resembles to some extent the unzipping of DNA under external force \cite{DNA1, DNA2}.

The TASEP -- \L{}ukasiewicz paths correspondence also elucidates the $\alpha \to \beta$ symmetry, i.e., the symmetry between the attractive field $U(x) = \delta_{1,x}\log \beta$, acting on all links of the \L{}ukasiewicz paths at a single point $x=1$, and the repulsive field $V(x) = x \log \alpha$ acting only on the end link of the \L{}ukasiewicz path but at any $x$. To the best of our knowledge, this rather nontrivial symmetry has never been discussed before.

\section{TASEP and HP from the underlying lattice gas}
\label{c}

\subsection{Viennot theorem}

In this section we have explained how the HP problem and the steady-state TASEP are connected (by virtue of the mapping described in the previous section) with the partition function of a one-dimensional gas of particles with hardcore interactions. This connection is based on a theorem first proved in \cite{viennot-rev}, which links the generating function of a heap of pieces with the generating function of a single layer of the heap. We start with stating the general formulation of the theorem, and then apply it for the particular case of the heap of square pieces with no common vertical sides.

Assume that $Z_N$ is a partition function of a heap constructed over some given graph $\cal G$ as a base, where the vertices of the graph $\cal G$ designate possible locations of the elementary pieces, and the edges of the graph designate the vertices which cannot be simultaneously occupied in a single layer (in our particular case the graph is just a chain of $n$ vertices). Let $\Xi (s)$ be the corresponding generating function (grand canonical partition function):
\be
\Xi(s) = \sum_{N=1}^{\infty} Z_N s^N \equiv \sum_{\text{allowed configurations}} s^{\# \text{~of pieces}},
\label{eq:34}
\ee
Define also the partition function $\Theta(k)$ of all possible distinct configurations of $k$ elementary pieces in a single layer, i.e. all possible subsets of $k$ vertices of $\cal G$, such that no edge has both its ends included into the subset, and the corresponding generating function
\be
\Omega(s) = 1 + \sum_{k=1}^{k_{\max}} \Theta(k) s^k,
\label{eq:35}
\ee
where $k_{\max}$ is the maximal possible size of such a subset. In this formulation $\Omega(s)$ is the partition function of a hard-core lattice gas on $\cal G$ with fugacity $s$.

\begin{figure}[ht]
\epsfig{file=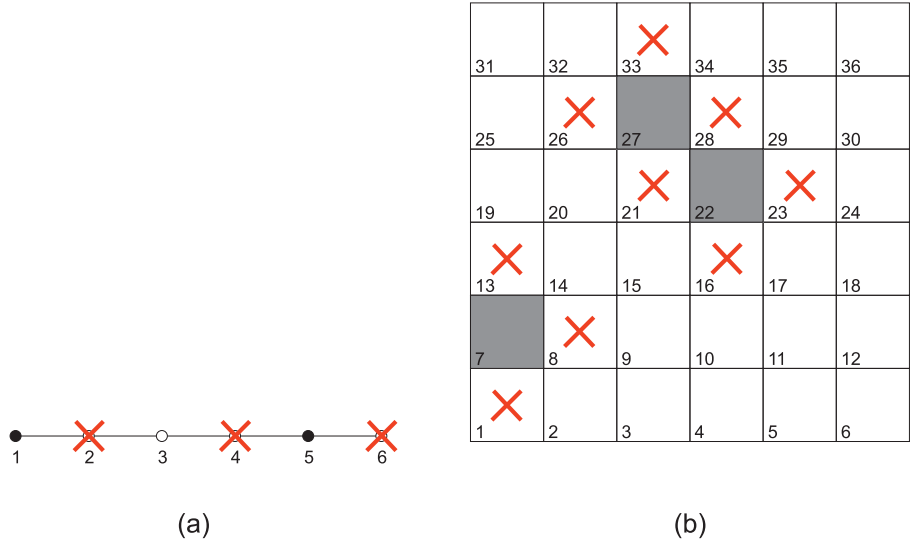, width=12cm}
\caption{Sample configurations of particles with hard-core interactions on a segment ($D=1$) (a) and on a square lattice ($D=2$) (b). Particles are denoted by filled elementary units (small circles for chain, squares for two-dimensional lattice), crosses mark positions which are forbidden for the particles.}
\label{f:06}
\end{figure}

Then the theorem \cite{viennot-rev} states that
\be
\Xi(s) = \frac{1}{\Omega(-s)}.
\label{eq:37}
\ee
For completeness, we present here a sketch of the proof. Consider the product $\Xi(s) \Omega(t)$, which enumerates configurations in the direct product of:
\begin{enumerate}
\item[(a)] The set of {\it all possible heaps}, enumeration of whose pieces is generated by $s$,
\item[(b)] The set of {\it all possible single layers}, enumeration of whose pieces is generated by $t$.
\end{enumerate}
For brevity, call the first set ``a heap of $s$-pieces'', and the second set -- ``a layer of $t$-pieces''. Consider now an element of the direct product (i.e. a pair of a heap and a layer), and put the heap on top of the layer, i.e. put the layer at the bottom floor, so that all $t$-pieces have vertical coordinate 0, and then put the heap on top of it (i.e. shift all vertical coordinates of the elements of the heap by 1). The resulting configuration is, generally speaking, not a heap of pieces itself: it is possible that some pieces of the $s$-heap are not supported from below by the elements of the $t$-layer. If it is the case, we allow such pieces to fall down to the underlying layer until no further rearrangements are possible. The generating function of all resulting structures can be written in a following way
\be
\Xi(s) \Omega(t) = \sum_{\alpha} t^{n_\alpha} \sum_{\beta}s^{n_\beta} F_{\alpha,\beta}(s),
\label{eq:38}
\ee
where $\alpha$ and $\beta$ enumerate all possible configurations of $t$- and $s$-pieces in the lowest layer (i.e., the combination of $t$-pieces being the original layer configuration, and combination of $s$-pieces no matter where they fell from the upper layer), $n_{\alpha,\beta}$ are the respective numbers of pieces in the lowest layer, and $F_{\alpha,\beta}(s)$ is the generating function of all heaps that can be placed on top of a fixed lowest layer configuration. Now, the crucial idea is that $F_{\alpha,\beta}(s)$ is a function of only the {\it total} configuration of the lowest layer, $\alpha \cup \beta$, and not of the way how the pieces are separated into $s$-type and $t$-type. Therefore,
\be
\Xi(s) \Omega(t) = \sum_{\alpha\cup\beta} F_{\alpha \cup \beta}(s) \sum_{\alpha} t^{n_\alpha} s^{n_\beta} = \sum_{\alpha\cup\beta} F_{\alpha \cup \beta}(s) (s+t)^{n_{\alpha \cup \beta}},
\label{eq:39}
\ee
where the first sum runs over all possible configurations of the lowest layer, and the second -- over all possible separations of lowest level pieces into $s$- and $t$-types. The last equation allows for the fact that each piece can be assigned to either $s$- or $t$-type independently of others. Note now, that \eq{eq:39} is radically simplified for $t = -s$. Indeed, only the term with $n_{\alpha \cup \beta}=0$ (i.e. which corresponds to an empty layer and an empty heap) survives, and therefore
\be
\Xi (s) \Omega(-s) = 1,
\label{eq:40}
\ee
completing the Viennot's theorem.

The function $Z_N$ can be obtained from $\Xi(s)$ in a standard way
\be
Z_N = \frac{1}{2\pi i} \oint \frac{\Xi(s)}{s^{N+1}}ds
\label{eq:41}
\ee
and, therefore, the growth rate, i.e., the leading large-$N$ asymptotics of the partition function \eq{Lambda-def} is controlled by the singularity of $\Xi(s)$ with the smallest absolute value. Taking into account \eq{eq:37} this means that
\be
\Lambda = -s_*^{-1},
\label{eq:42}
\ee
where $s_*$ is negative and the smallest in absolute value number (among all zeros and non-pole singularities of the generating polynomial $\Omega(s))$.

By virtue of \eq{eq:40}, the combinatorics of (D+1)-dimensional HPs can be reformulated as a problem of calculating the grand canonical partition function of a D-dimensional ``hard-square lattice gas'', which in turn can be thought of as a D-dimensional Ising model with finite magnetic field in the limit of strong anti-ferromagnetic coupling.

The negative Yang-Lee zero closest to the origin is associated with a point where the thermodynamic functions of a hard-core gas in the thermodynamic limit are known to exhibit a ``non-physical'' singularity on the negative real fugacity axis \cite{Groeneveld62,Gaunt65,Gaunt69,Assis13}. This point sometimes is called ``the Lee-Yang critical point'' \cite{Bouttier02}. This is a remarkably general feature of systems with repulsive interactions, which have pressure function singularities for complex values of the chemical potential (see, e.g. \cite{Taradiy19}). It was argued that systems with repulsive interactions possess universal properties associated with the dominant singularity of the Mayer fugacity series \cite{Poland84,Baram87}. Subsequently, it was shown that indeed this singularity can be identified with the Yang-Lee edge singularity \cite{Lai95,Todo99}.

\subsection{Generating function of a 1D hard core lattice gas with an adsorbing site}

Viennot theorem, as described above, is formulated for heaps with identical layers and is directly applicable to the unweighted and unrestricted heaps. In the terminology of section III the statement of the theorem can be written as follows
\be
\Xi_n^\text{HP} (s)= \sum_{N=0}^{\infty} Z_{N,n}^{\text{HP}} (\beta) s^N = \Omega_n^{-1}(-s,\beta),
\label{viennotHP}
\ee
where $Z_{N,n}^{\text{HP}}(\beta)$ is the total partition function of all heaps in the $n\times \infty$ box with adsorbing right wall,
\be
Z_{N,n}^{\text{HP}}(\beta)=\langle 1,1,\dots,1|T_{\beta}^N(n)| 1,1,\dots,1\rangle,
\label{ZHP}
\ee
$T_{\beta}(n)$ is given by \eq{eq:19}, and $\Omega_n(s)$ is the grand partition function of the corresponding one-layer problem, i.e., a 1D lattice gas with hard-core interactions (two pieces cannot occupy adjacent sites) and statistical weight $\beta^{-1}$ associated with the leftmost site ($n$ here is the number of accessible lattice sites). For $\beta =1$ this partition function obeys the equation
\be
\Omega_{n+2}(s,1) = \Omega_{n+1}(s,1) + s \Omega_{n}(s,1),\,\,
\Omega_0 =1,\,\,
\Omega_1= 1+s.
\label{eq:Q}
\ee
Solving \eq{eq:Q} similarly to \eq{eq:25}, we get
\be
\Omega_n(s,1)=\frac{1+s-q_2}{q_1-q_2} q_1^{n}-\frac{1+s-q_1}{q_1-q_2} q_2^{n};
\qquad q_{1,2}=\frac{1\pm \sqrt{1+4s}}{2}.
\label{eq:Zn}
\ee
In the general case $\Omega_n(s,\beta)$ satisfies the recursion
\be
\Omega_n(s,\beta) = \Omega_{n-1}(s,1) + \frac{s}{\beta} \Omega_{n-2}(s,1).
\label{Zbeta}
\ee
Substituting \eq{eq:Zn} into \eq{Zbeta} and collecting the terms leads to the following expression for $\Omega_n(s,\beta)$:
\be
\Omega_n(s,\beta) = \frac{1}{\sqrt{1+4s}}\left(\left(q_1+\frac{s}{\beta}\right) q_1^n - \left(q_2+\frac{s}{\beta}\right) q_2^n\right); \qquad q_{1,2} = \frac{1\pm \sqrt{1+4s}}{2}.
\label{Zbeta2}
\ee
Together with \eq{viennotHP} this allows to recover the grand partition function of a heap of pieces
\be
\Xi_n^\text{HP}(s,\beta) = \Omega_n^{-1}(-s,\beta).
\label{HP_Viennot}
\ee
Note that despite the formal presence of square roots in \eq{Zbeta2}, $\Omega_n(s,\beta)$ is a polynomial of order $\lfloor(n+1)/2\rfloor$ in $s$, and thus the growth rate of the HP is controlled by its largest negative zero. Thus, the growth rate of a HP, given by \eq{eq:09} in the case of $\beta = 1$, is governed by the Lee-Yang zero of the partition function of the corresponding one-dimensional gas. To check this correspondence, we invite the reader to re-derive \eq{eq:09} directly from \eq{Zbeta2}.

Now, the mapping described in section III links the weights of TASEP configurations in the stationary state with the enumeration of {\it weighted pyramids} in the HP problem. Weighted pyramids are not heaps of identical layers, and thus Viennot theorem is not directly applicable to them. However, the partition function of pyramids $\Xi_n (s,\alpha, \beta)$, given by \eq{Xi}, which converges to the generating function of stationary TASEP in the large $n$ limit, is very similar to the partition function of all HP configurations \eq{viennotHP}--\eq{ZHP}, essentially they correspond to different matrix elements of the same matrix $\left(s T_{\beta}(n) - I \right)^{-1}$ (see \eq{eq:21}). It is therefore not surprising that the determinant of this matrix is
\be
\det\left (s T_{\beta}(n) - I \right) = s^n u_n(s,\beta) = \Omega_n(-s,\beta).
\label{viennot1}
\ee

Recall that the partition function, $W_n(s,k)$, of all the pyramids which have the last piece at position $k$, is a quotient of such determinants (see \eq{eq:22}, \eq{eq:31})
\be
W_n(s,k)=\frac{v_{n,k}}{u_n} = \frac{(-1)^k}{s} \frac{u_{n-k}(\beta = 1)}{u_n(\beta)} = (-1)^k s^{k-1} \frac{ \Omega_{n-k}(-s,1)}{ \Omega_n(-s,\beta)}
\ee

Thus, the partition function of weighted pyramids \eq{Xi_n} can be written in terms of the partition function of the 1D ideal gas with hard-core interaction $\Omega_n(-s,\beta)$. Indeed,
\be
\Xi_n (s,\alpha, \beta) =- \left(-\frac{s}{\alpha}\right)^{n-1} \frac{1}{\Omega_n(-s,\beta)} \sum_{k=1}^n \left(-\frac{\alpha}{s}\right)^{n-k} \Omega_{n-k}(-s,1) = - \left(-\frac{s}{\alpha}\right)^{n-1} \frac{\tilde{\Omega}_n(-s,-\alpha/s)}{\Omega_n(-s,\beta)},
\label{eq:Xi}
\ee
where we introduced the partial generating function
\be
\tilde{\Omega}_n(s,t) = \sum_{m=0}^{n-1} \Omega_{m}(s,1) t^m.
\label{eq:Omega}
\ee
Note that $\tilde{\Omega}_n(-s,-\alpha/s)$ is, with respect to $1/s$, a polynomial of power $n-1$, so $\Xi_n (s,\alpha, \beta)$ converges to a finite value for $s \to 0$. It is possible to take the large $n$ limit of this expression explicitly and get back to the formula \eq{TASEP_Xi} for the grand partition function of the stationary TASEP, $\Xi (s,\alpha, \beta) = \lim_{n \to \infty} \Xi_n (s,\alpha, \beta)$ (not once again that the $\alpha \leftrightarrow \beta$ symmetry appears only in the limiting formula). This establishes the desired connection between the TASEP problem with free boundary conditions and the partition function of a 1D hard-core lattice gas on a strip with adsorbing boundary.

\section{Discussion}
\label{d}

In this paper we studied the multiple connections among basic classical models of statistical physics: (i) the 1D lattice gas with hard-core interactions, (ii) the 1D TASEP with open boundary conditions, (iii) the problem of (1+1)D heaps enumerations of square pieces with hard-core repulsion in the horizontal direction, and (iv) an ideal (1+1)D polymer chain represented by a \L{}ukasiewicz path. By exploiting various mappings between these problems, and the X. Viennot theorem connecting partition functions of a heap of pieces and that of a single layer of pieces, we were able to show eventually that the partition function of the steady-state TASEP with open boundary conditions can be expressed in terms of a quotient of partition functions \eq{eq:Xi}--\eq{eq:Omega} of a one-dimensional hard-core lattice gas with an adsorbing site at the boundary and negative fugacity.

Although all the used individual mappings were already present in the literature, this final result has not, to the best of our knowledge, been reported before. It provides, in our opinion, an important advancement of connections between sonsidered statistical systems. Another interesting and previously unknown mapping is the connection between the three phases in the steady state TASEP with open boundary conditions, and the three states of an ideal polymer chain on a half-line with an adsorbing wall and an external field acting on the end link. The latter connection highlights a non-trivial hidden symmetry between adsorbing potential acting on all the links in the vicinity of the wall, and a repulsive field, which is independent of the distance to the wall, but acts only onto the end monomer.

Notably, Viennot's theorem can be exploited further to establish connections between the Yang-Lee zeros of the D-dimensional lattice gas with excluded volume interactions and the enumeration of a (D+1)-dimensional heap of pieces (see \cite{npt20}). This is a nice example of a problem in which Yang-Lee zeros have a direct physical meaning.

\begin{acknowledgments}
We are grateful to S. Redner and D. Dhar for many illuminating discussions. The work of S.N. is supported by the BASIS Foundation in frameworks of the grant 19-1-1-48-1.
\end{acknowledgments}

\begin{appendix}

\section{Group-theoretical approach to the counting of HP}

Here we provide a group-theoretical interpretation of an enumeration of heaps introduced in section IIIA. In \fig{f:01}a,b two particular realizations of a heap of $N=11$ particles are shown. These two examples correspond to different sequential depositions of pieces but the resulting heaps are geometrically the same. In order to define the equivalence of heaps, it is instructive to use the following auxiliary construction. Let $F_n$ be the group (called in \cite{vershik} the ``locally-free group'') defined on a set of $n$ generators $\{g_1, g_2,..., g_n\}$ which obey the following commutation relations
\be
g_k g_m = g_m g_k, \qquad |k-m| \ge 2
\label{eq:01}
\ee
The group element of length $N$ is an arbitrary $N$-letter word written in terms of generators $\{g_1, g_2,..., g_n\}$. Consider the positive semigroup, $F_n^+$ of this group, i.e. exclude all the words of the group which involve inverse generators $g_j^{-1}$ ($1\le j\le n$). Now, there exists a one-to-one correspondence between configurations of HP and equivalence classes of words in the semigroup $F_n^+$.

In \cite{vershik} it has been shown that the partition function of an $N$-particle heap of pieces in a 2D bounding box of $n$ columns coincides with the partition function of a special $N$-step Markov chain on $F_n^+$. Namely, any configuration of HP can be bijectively associated with a class of equivalent words in $F_n^+$. Each equivalence class is represented by a {\it unique} word written in a normally ordered sequence of letters-generators of $F_n^+$. Normal ordering means that generators with smaller indices are pushed to the left when it is allowed by the commutation relations \eq{eq:01}. Consequently, the word $w=g_{s_1} g_{s_2}\ldots g_{s_N}$ is in normal form if and only if the indices $s_1,...,s_N$ satisfy the following conditions:
\be
\begin{array}{ll}
\mbox{(a)} & \mbox{If $s_N=x$ ($1\le x\le n-1$) then $s_{N-1}\in\{1,2,...,x+1\}$} \medskip \\ \mbox{(b)} & \mbox{If $s_N=n$ then $s_{N-1}\in\{1,2,..., n\}$}
\end{array}
\label{eq:02}
\ee
Let us demonstrate how \eq{eq:02} works for heaps shown in \fig{f:01}a,b. Denote by $g_j$ ($1\leq j\leq n$) the deposition of a piece into the column $j$ ($1\le j \le 5$). The heap shown in \fig{f:01}a is constructed by a consecutive deposition of $g_1$, then $g_2$ etc, so that the entire heap is encoded by a 11-letter word $W_a$:
$$
W_a=g_3\,g_1\,g_5\,g_1\,g_2\,g_4\,g_5\,g_2\,g_3\,g_4\,g_2\,g_1
$$
The heap shown in \fig{f:01}b is obtained by dropping pieces in a different order. The corresponding word $W_b$ spells:
$$
W_b=g_5\,g_3\,g_4\,g_5\,g_1\,g_1\,g_2\,g_2\,g_3\,g_1\,g_1\,g_4
$$
Although the words $W_a$ and $W_b$ are different, they encode the same configuration of pieces. To establish the bijection between a heap and a word, we write down words $W_a$ and $W_b$ in a standard i.e. ``normally ordered'' form. Namely, we push the generators with smaller indices in the words $W_a$ and $W_b$ as left as possible when that is consistent with the commutation relations \eq{eq:02}. One can straightforwardly verify that after such a reordering, both words $W_a$ and $W_b$ get transformed into the word $\overrightarrow{W}$:
\be
\overrightarrow{W}=g_1\,g_1\,g_3\,g_2\,g_2\,g_5\,g_4\,g_3\,g_2\,g_1\,g_5\,g_4.
\label{eq:03}
\ee
Thus, any $N$-site heap in a bounding box of $n$ columns can be uniquely represented by a $N$--letter word ``spelled'' by the generators of $F_n^+$ in a normal order.

\section{\L{}ukasiewicz paths, Dyck paths and Young tableaux}
\label{sect:luk}

\begin{figure}[h]
\epsfig{file=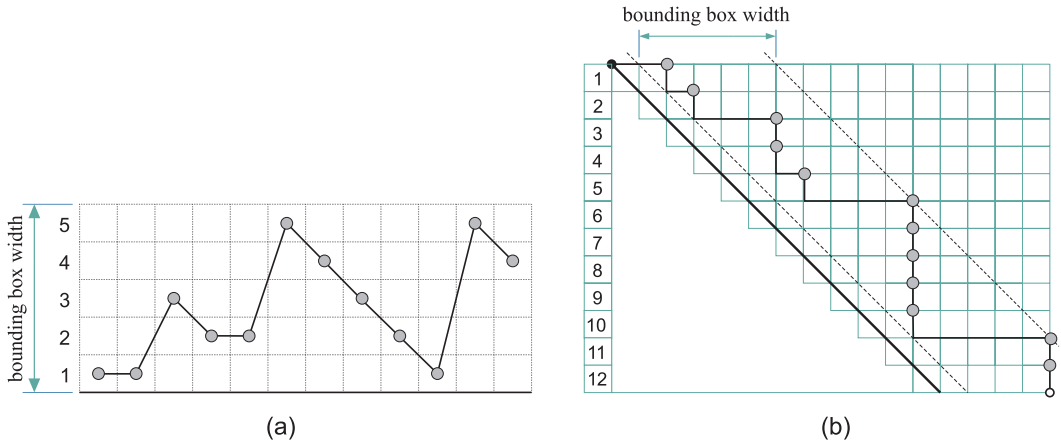, width=14cm}
\caption{(a) The \L{}ukasiewicz path corresponding to the normally ordered word $g_1\, g_1\, g_3\, g_2\, g_2\, g_5\, g_4\, g_3\, g_2\, g_1\, g_5\, g_4$ and the heap shown in \fig{f:02}; (b) the corresponding path in skewed coordinates (see text), which can be interpreted as a Dyck path or a Young tableaux.}
\label{f:04}
\end{figure}

Here we elucidate the connection between \L{}ukasiewicz paths used for the enumeration of heaps of pieces in the main text, the more conventional Dyck paths, and the Young tableaux (similar mapping has been presented in the literature, e.g., in \cite{prellberg}). Consider a \L{}ukasiewicz path, i.e. a discrete random walk in (1+1)D consisting of steps $(1,x)$ where $x$ is integer and $x \geq - 1$, and redraw the walk as follows. Keep all the down steps of the original \L{}ukasiewicz path, replace each horizontal step with a sequence of down and up steps, and each up step of length $x$ replace with a sequence of a single down step and $x+1$ up steps. The resulting trajectory, as shown in \fig{f:04} in the coordinates rotated by the angle $\pi/4$, is nothing but a ordinary (1+1)D Dyck random walk, consisting of up and down steps wit respect to the diagonal of a rotated square. Clearly, this procedure creates a bijection between \L{}ukasiewicz and Dyck paths. Note, however, that the length of the walk is not conserved by this procedure: a \L{}ukasiewicz path of length $N$ starting at 0 and ending in a point with vertical coordinate $x$ gets transformed into a Dyck path of length $2N+x$ ending in a point with vertical coordinate $x$. In particular, \L{}ukasiewicz excursions of length $N$ (i.e., paths starting and ending at 0) are thus mapped onto Dyck excursions of length $2N$. 

From the other hand, the trajectory shown in \fig{f:04} can be viewed as a Young tableau (YT). By definition, the enveloping shape of the YT should not be concave (in the so-called ``French notation'' of YT). Let us enumerate rows above the diagonal in \fig{f:04}b upside down and let $x_t$ be the length of the $t$th row. By definition of the YT, the length $x_{t+1}$ of the row $t+1$ can take any value from the set $\{x_t-1, x_t, x_t+1,x_t+2,..., n\}$, where $x_t$ is the length of the nearest upper row, $t$. Comparing with the definition of a \L{}ukasiewicz path, we conclude that it is a particular realization of a Young tableau, being a rephrasing of the standard representation shown in \fig{f:04}b.

The shape of the Young tableau in \fig{f:04}b and the realization of the particular \L{}ukasiewicz path shown in \fig{f:04}a) is uniquely encoded by a sequence of generators $g_i$. The YT in \fig{f:04}a is represented by the word $W_{YT}=g_1\, g_1\, g_3\, g_2 \, g_2\, g_5\, g_4\, g_3\, g_2\, g_1\, g_5\, g_4$ which is nothing else but the ``Mikado ordered'' word written in terms of generators of the locally free semigroup $F^+$, defined in Appendix A.



\end{appendix}

\end{document}